\documentclass{article}%
\usepackage{amsmath}
\usepackage{amsfonts}
\usepackage{amssymb}
\usepackage{graphicx}%
\providecommand{\U}[1]{\protect\rule{.1in}{.1in}}

\begin{document}

\title{Entropic Dynamics: from Entropy and Information Geometry to Hamiltonians and
Quantum Mechanics\thanks{Presented at MaxEnt 2014, the 34th International
Workshop on Bayesian Inference and Maximum Entropy Methods in Science and
Engineering (September 21--26, 2014, Amboise, France). }}
\author{Ariel Caticha, Daniel Bartolomeo\\{\small Physics Department, University at Albany-SUNY, Albany, NY 12222, USA.}
\and Marcel Reginatto\\{\small Physicalisch-Technische Bundesanstalt, 38116 Braunschweig, Germany}}
\date{}
\maketitle

\begin{abstract}
Entropic Dynamics is a framework in which quantum theory is derived as an
application of entropic methods of inference. There is no underlying action
principle. Instead, the dynamics is driven by entropy subject to the
appropriate constraints. In this paper we show how a Hamiltonian dynamics
arises as a type of non-dissipative entropic dynamics. We also show that the
particular form of the \textquotedblleft quantum potential\textquotedblright%
\ that leads to the Schr\"{o}dinger equation follows naturally from
information geometry.

\end{abstract}

\section{Introduction}

In the standard view quantum theory (QT) is a type of mechanics and it is
natural to postulate that its dynamical laws are given by an action principle.
In contrast, Entropic Dynamics (ED) views quantum theory as an application of
entropic methods of inference and there is no underlying action principle. The
dynamics is generated by continuously maximizing an entropy as constrained by
the appropriate relevant information --- it is through these constraints that
the \textquotedblleft physics\textquotedblright\ is introduced. \cite{Caticha
2010}\cite{Caticha 2014} The ED approach allows a fresh perspective on
familiar notions such as time and mass and on long-standing conceptual
difficulties, such as indeterminism and the problem of measurement.

The early formulations of ED involved assumptions that were justified only by
their pragmatic success --- they led to the right answers. For example, use
was made of auxiliary variables the physical interpretation of which remained
obscure and there were further assumptions about the configuration space
metric and the form of the quantum potential. In \cite{Caticha 2014} it was
shown that the auxiliary variables were in fact unnecessary and could be eliminated.

In this paper the derivation of QT as a form of entropic dynamics is further
strengthened by establishing its relation to information geometry and to
Hamiltonian dynamics. We show that a non-dissipative entropic dynamics
naturally leads to a Hamiltonian formalism including an action principle. The
metric of the $N$-particle configuration space does not need to be postulated;
we derive it from information geometry and show that it coincides with the
mass tensor. Finally, the particular form of Hamiltonian that leads to QT
requires a so-called \textquotedblleft quantum potential\textquotedblright%
\ which, we show, is a natural construct within information
geometry.\footnote{Additional references to entropic dynamics and to other
information-based approaches to quantum theory including the relation to
information geometry are given in \cite{Caticha 2010}\cite{Caticha
2014}\cite{Reginatto 2013}.}

\section{Entropic Dynamics}

In order to formulate QT as an example of entropic inference\footnote{For an
overview of Bayesian and entropic inference and further references see
\cite{Caticha 2012}.} we must identify the microstates that are the subject of
our inference, we must identify prior probabilities, and we must identify
those constraints that represent the information that is relevant to our
problem. First the microstates: We consider $N$ particles living in flat
Euclidean space $\mathcal{X}$ with metric $\delta_{ab}$. The particles have
definite positions $x_{n}^{a}$ and it is their unknown values that we wish to
infer.\footnote{In this work ED is developed as a model for the quantum
mechanics of particles. The same framework can be deployed to construct models
for the quantum mechanics of fields, in which case it is the fields that are
objectively \textquotedblleft real\textquotedblright\ and have well-defined
albeit unknown values.\cite{Ipek Caticha 2014}} (The index $n$ $=1\ldots N$
denotes the particle and $a=1,2,3$ the spatial coordinate.) For $N$ particles
the configuration space $\mathcal{X}_{N}=\mathcal{X}\times\ldots
\times\mathcal{X}$.

The basic dynamical assumption is that motion is continuous, that is, large
displacements are possible but only as a result of the accumulation of many
small steps. We do not explain why motion happens but, given the information
that it does, our task is to venture a guess about what to expect. Thus, we
first consider a single short step and later we determine how the accumulation
of short steps yields a large displacement.

The first goal is to find the transition probability density $P(x^{\prime}|x)$
for a single short step from a given initial $x\in\mathcal{X}_{N}$ to an
unknown $x^{\prime}\in\mathcal{X}_{N}$. The starting point is a prior
transition probability $Q(x^{\prime}|x)$ that expresses our a priori knowledge
about which $x^{\prime}$ to expect \emph{before} any information about the
expected step is taken into account. Next, the physically relevant information
about the step is expressed in the form of constraints that $P(x^{\prime}|x)$
must satisfy --- this is the stage in which the physics is introduced.
Finally, the method of maximum entropy is used to update from the prior
probability $Q(x^{\prime}|x)$ to the desired posterior probability
$P(x^{\prime}|x)$. More specifically, to find $P(x^{\prime}|x)$ we maximize
the (relative) entropy,
\begin{equation}
\mathcal{S}[P,Q]=-\int d^{3N}x^{\prime}\,P(x^{\prime}|x)\log\frac{P(x^{\prime
}|x)}{Q(x^{\prime}|x)}~. \label{Sppi}%
\end{equation}
subject to the physically relevant constraints.

We adopt a prior $Q(x^{\prime}|x)$ that represents a state of extreme
ignorance: knowledge of the initial position $x$ tells us nothing about
$x^{\prime}$. Such ignorance is expressed by assuming that $Q(x^{\prime
}|x)d^{3N}x^{\prime}$ is proportional to the volume element in $\mathcal{X}%
_{N}$. Since $\mathcal{X}_{N}$ is flat and the proportionality constant has no
effect on the entropy maximization we can set $Q(x^{\prime}|x)=1$%
.\footnote{Uniform non-normalizable priors are mathematically problematic.
This is a mild annoyance that is evaded by adopting a normalizable prior such
as a Gaussian centered at $x$ with a sufficiently large standard deviation.}

Next we introduce some information about the motion. The first piece of
information is that motion is continuous---it occurs as a succession of
infinitesimally short steps. Each individual particle $n$ will take a short
step from $x_{n}^{a}$ to $x_{n}^{\prime a}=x_{n}^{a}+\Delta x_{n}^{a}$ and we
require that the expected squared displacement,
\begin{equation}
\langle\Delta x_{n}^{a}\Delta x_{n}^{b}\rangle\delta_{ab}=\kappa_{n}%
~,\qquad(n=1\ldots N)~ \label{kappa n}%
\end{equation}
be some small value $\kappa_{n}$. For infinitesimally short steps we will
eventually take the limit $\kappa_{n}\rightarrow0$. To reflect the
translational symmetry of $\mathcal{X}$ we will assume each $\kappa_{n}$ to be
independent of $x$. However, in order to account for differences among
non-identical particles we allow $\kappa_{n}$ to depend on the particle index
$n$. The constraint (\ref{kappa n}) leads to a completely isotropic diffusion.
Directionality is introduced by assuming the existence of a \textquotedblleft
potential\textquotedblright\ $\phi(x)$ and imposing a constraint on the
expected displacement $\left\langle \Delta x\right\rangle $ along the gradient
of $\phi$,\footnote{Elsewhere, in the context of particles with spin, we will
see that the potential $\phi(x)$ can be given a natural geometric
interpretation as an angular variable. Its integral over any closed loop is $%
{\displaystyle\oint}
d\phi=2\pi n$ where $n$ is an integer.}
\begin{equation}
\langle\Delta x^{A}\rangle\partial_{A}\phi=\sum\limits_{n=1}^{N}\left\langle
\Delta x_{n}^{a}\right\rangle \frac{\partial\phi}{\partial x_{n}^{a}}%
=\kappa^{\prime}~, \label{kappa prime}%
\end{equation}
where $\partial_{A}=\partial/\partial x^{A}=\partial/\partial x_{n}^{a}$
(capitalized indices such as $A=(n,a)$ denote both the particle index and its
spatial coordinate). $\kappa^{\prime}$ is another small but for now
unspecified position-independent constant.

Varying $P(x^{\prime}|x)$ to maximize $\mathcal{S}[P,Q]$ in (\ref{Sppi})
subject to the $N+2$ constraints (\ref{kappa n}), (\ref{kappa prime}) and
normalization gives
\begin{equation}
P(x^{\prime}|x)=\frac{1}{\zeta}\exp[-\sum_{n}(\frac{1}{2}\alpha_{n}\,\Delta
x_{n}^{a}\Delta x_{n}^{b}\delta_{ab}-\alpha^{\prime}\Delta x_{n}^{a}%
\frac{\partial\phi}{\partial x_{n}^{a}})]~, \label{Prob xp/x a}%
\end{equation}
where $\zeta=\zeta(x,\alpha_{n},\alpha^{\prime})$ is a normalization constant
and $\alpha_{n}$ and $\alpha^{\prime}$ are Lagrange multipliers. Since both
the function $\phi$ and the constant $\kappa^{\prime}$ are so far unspecified
we can, without loss of generality, absorb $\alpha^{\prime}$ into $\phi$ which
amounts to setting $\alpha^{\prime}=1$. The distribution $P(x^{\prime}|x)$ is
Gaussian and is conveniently rewritten as
\begin{equation}
P(x^{\prime}|x)=\frac{1}{Z}\exp[-\frac{1}{2}\sum_{n}\alpha_{n}\,\delta
_{ab}(\Delta x_{n}^{a}-\langle\Delta x_{n}^{a}\rangle)(\Delta x_{n}%
^{b}-\langle\Delta x_{n}^{b}\rangle)]~, \label{Prob xp/x b}%
\end{equation}
where $Z$ is a new normalization constant. A generic displacement $\Delta
x_{n}^{a}=x_{n}^{\prime a}-x_{n}^{a}$ can be expressed as an expected drift
plus a fluctuation, $\Delta x_{n}^{a}=\left\langle \Delta x_{n}^{a}%
\right\rangle +\Delta w_{n}^{a}\,$, where
\begin{equation}
\left\langle \Delta x_{n}^{a}\right\rangle =\frac{1}{\alpha_{n}}\delta
^{ab}\frac{\partial\phi}{\partial x_{n}^{b}}~, \label{ED drift}%
\end{equation}%
\begin{equation}
\left\langle \Delta w_{n}^{a}\right\rangle =0\quad\text{and}\quad\langle\Delta
w_{n}^{a}\Delta w_{n}^{b}\rangle=\frac{1}{\alpha_{n}}\delta^{ab}~.
\label{ED fluctuations}%
\end{equation}
For very short steps, as $\alpha\rightarrow\infty$, the fluctuations become
dominant: the drift is $\Delta\bar{x}_{n}\sim\alpha_{n}^{-1}$ while $\Delta
w_{n}\sim\alpha_{n}^{-1/2}$. This implies that, as in Brownian motion, the
trajectory is continuous but not differentiable. In the ED approach a particle
has a definite position but its velocity, the tangent to the trajectory, is
completely undefined.

\section{Entropic time}

The foundation of all notions of time is dynamics. In ED time is introduced as
a book-keeping device to keep track to the accumulation of small changes. As
discussed in \cite{Caticha 2010}\cite{Caticha 2012} this involves introducing
a notion of instants that are ordered, and defining the interval or duration
between them. The idea is that if $\rho(x,t)$ refers to a probability
distribution at a given instant, which we label $t$, then entropic time is
constructed by defining the \emph{next} instant, labelled $t^{\prime}$, in
terms of a new distribution
\begin{equation}
\rho(x^{\prime},t^{\prime})=\int d^{3}x\,P(x^{\prime}|x)\rho(x,t)~,
\label{CK b}%
\end{equation}
where the transition probability for infinitesimally short steps is
$P(x^{\prime}|x)$ in eq.(\ref{Prob xp/x b}). The iteration of this process
defines the dynamics: entropic time is constructed instant by instant:
$\rho_{t^{\prime}}$ is constructed from $\rho_{t}$, $\rho_{t^{\prime\prime}}$
is constructed from $\rho_{t^{\prime}}$, and so on. \emph{\ }

Having introduced the notion of successive instants we now have to specify the
interval $\Delta t$ between them. This amounts to specifying the multipliers
$\alpha_{n}(x,t)$ in terms of $\Delta t$.

\emph{Time is defined so that motion looks simple.} For large $\alpha_{n}$ the
dynamics is dominated by the f{}luctuations $\Delta w_{n}$. In order that the
f{}luctuations $\left\langle \Delta w_{n}^{a}\Delta w_{n}^{b}\right\rangle $
ref{}lect the symmetry of translations in space and time --- a Newtonian time
that flows \textquotedblleft equably\ everywhere and
everywhen\textquotedblright\ --- we choose $\alpha_{n}$ to be independent of
$x$ and $t$,
\begin{equation}
\alpha_{n}=\frac{m_{n}}{\eta\Delta t}~. \label{alpha n}%
\end{equation}
The $m_{n}$ are particle-specific constants, which will eventually be
identified as particle masses, and $\eta$ is a particle-independent constant
that fixes the units of the $m_{n}$s relative to the units of time and will
eventually (after regraduation) be identified as $\hbar$.

\section{The information metric of configuration space}

To each point $x\in\mathcal{X}_{N}$ we can associate a probability
distribution $P(x^{\prime}|x)$. Thus, the configuration space $\mathcal{X}%
_{N}$ is a statistical manifold. Up to an arbitrary global scale factor its
geometry is uniquely determined by the information metric,%
\begin{equation}
\gamma_{AB}=C\int d^{3N}x^{\prime}\,P(x^{\prime}|x)\frac{\partial\log
P(x^{\prime}|x)}{\partial x^{A}}\frac{\partial\log P(x^{\prime}|x)}{\partial
x^{B}}~, \label{gamma C}%
\end{equation}
where $C$ is an arbitrary positive constant. (See \emph{e.g.}, \cite{Caticha
2012}.) For short steps ($\alpha_{n}\rightarrow\infty$) a straightforward
substitution of eq.(\ref{Prob xp/x b}) using eq.(\ref{alpha n}) yields
\begin{equation}
\gamma_{AB}=\frac{Cm_{n}}{\eta\Delta t}\delta_{nn^{\prime}}\,\delta_{ab}%
=\frac{Cm_{n}}{\eta\Delta t}\delta_{AB}~. \label{gamma AB}%
\end{equation}
We see that if $\Delta t\rightarrow0$ then $\gamma_{AB}\rightarrow\infty$. For
smaller $\Delta t$ the distributions $P(x^{\prime}|x)$ and $P(x^{\prime
}|x+\Delta x)$ become more sharply peaked and it is easier to distinguish one
from the other which translates into a greater information distance. In order
to define a distance that remains meaningful for arbitrarily small $\Delta t$
it is convenient to choose $C\propto\Delta t$. In what follows the metric
tensor will always appear in combinations such as $\gamma_{AB}\Delta t/C$. It
is therefore convenient to define the \textquotedblleft mass\textquotedblright%
\ tensor,
\begin{equation}
m_{AB}=\frac{\eta\Delta t}{C}\gamma_{AB}=m_{n}\delta_{AB}~,
\end{equation}
and its inverse, the \textquotedblleft diffusion\textquotedblright\ tensor,
\begin{equation}
m^{AB}=\frac{C}{\eta\Delta t}\gamma^{AB}=\frac{1}{m_{n}}\delta^{AB}~.
\end{equation}

With the choice of the multipliers $\alpha_{n}$ in (\ref{alpha n}) the
dynamics is indeed simple: $P(x^{\prime}|x)$ in\ (\ref{Prob xp/x b}) is a
standard Wiener process. The displacement is
\begin{equation}
\Delta x^{A}=b^{A}\Delta t+\Delta w^{A}~, \label{Delta x}%
\end{equation}
where $b^{A}(x)$ is the drift velocity,
\begin{equation}
\langle\Delta x^{A}\rangle=b^{A}\Delta t\quad\text{with}\quad b^{A}=\frac
{\eta}{m_{n}}\delta^{AB}\partial_{B}\phi=\eta m^{AB}\partial_{B}\phi~,
\label{drift velocity}%
\end{equation}
and the f{}luctuations $\Delta w^{A}$ satisfy,
\begin{equation}
\langle\Delta w^{A}\rangle=0\quad\text{and}\quad\langle\Delta w^{A}\Delta
w^{B}\rangle=\frac{\eta}{m_{n}}\delta^{AB}\Delta t=\eta m^{AB}\Delta t~.
\label{fluc}%
\end{equation}
Two remarks are in order: one on the nature of clocks and another on the
nature of mass.

\noindent\textbf{On clocks: }Time is defined so that motion looks simple. In
Newtonian mechanics the prototype of a clock is the free particle and time is
defined so that the free particle moves equal distances in equal times. In ED
the prototype of a clock is a free particle too --- for sufficiently short
times all particles are free --- and time is defined so that the particle
undergoes equal fluctuations in equal times.

\noindent\textbf{On mass:} The particle-specific constants $m_{n}$ will, in
due course, be called `mass' and eq.(\ref{fluc}) provides the interpretation:
mass is an inverse measure of fluctuations. Thus, up to overall constants the
metric of configuration space is the mass tensor and its inverse is the
diffusion tensor. In standard QM there are two mysteries: \textquotedblleft
Why quantum fluctuations?\textquotedblright\ and \textquotedblleft What is
mass?\textquotedblright. ED offers some progress in this matter: we do not
have two mysteries but just one. Fluctuations and mass are two sides of the
same coin.

\section{Accumulating changes: the Fokker-Planck equation}

Equation (\ref{CK b}) is an integral equation for the evolution of $\rho
(x,t)$. As is well known (see \emph{e.g.}, \cite{Caticha 2012}) it can be
written in differential form as a Fokker-Planck (FP) equation,
\begin{equation}
\partial_{t}\rho=-\partial_{A}\left(  b^{A}\rho\right)  +\frac{1}{2}\eta
m^{AB}\partial_{A}\partial_{B}\rho~. \label{FP a}%
\end{equation}
which can be rewritten as a continuity equation,
\begin{equation}
\partial_{t}\rho=-\partial_{A}\left(  \rho v^{A}\right)  ~. \label{FP b}%
\end{equation}
where $v^{A}$ is the velocity of the probability flow or \emph{current
velocity},
\begin{equation}
v^{A}=b^{A}+u^{A}\quad\text{and}\quad u^{A}=-\eta m^{AB}\partial_{B}\log
\rho^{1/2}~
\end{equation}
is the \emph{osmotic velocity}, which represents the tendency for probability
to flow down the density gradient. Since both $b^{A}$ and $u^{A}$ are
gradients, it follows that the current velocity is a gradient too,%
\begin{equation}
v^{A}=m^{AB}\partial_{B}\Phi\quad\text{where}\quad\frac{\Phi}{\eta}=\phi
-\log\rho^{1/2}~. \label{curr}%
\end{equation}
The FP equation
\begin{equation}
\partial_{t}\rho=-\partial_{A}\left(  \rho m^{AB}\partial_{B}\Phi\right)  ~,
\label{FP c}%
\end{equation}
can be conveniently rewritten in the alternative form
\begin{equation}
\partial_{t}\rho=\frac{\delta H}{\delta\Phi}~, \label{Hamilton a}%
\end{equation}
for some suitably chosen functional $H[\rho,\Phi]$. It is easy to check that
the appropriate functional $H$ is%
\begin{equation}
H[\rho,\Phi]=\int dx\,\frac{1}{2}\rho m^{AB}\partial_{A}\Phi\partial_{B}%
\Phi+F[\rho]~, \label{Hamiltonian a}%
\end{equation}
where $F[\rho]$ is some unspecified functional of $\rho$. In what follows we
will assume that $F=F[\rho]$ rather than the more general $F[\rho;t]$. It is
worth emphasizing that eqs.(\ref{FP b}), (\ref{FP c}), and (\ref{Hamilton a})
do not reflect new dynamical principles but are merely different ways to
rewrite the very same entropic dynamics already expressed by the FP
eq.(\ref{FP a}).

With these results ED reaches a certain level of completion: We figured out
what small changes to expect and time was introduced to keep track of how
these small changes accumulate; the net result is a standard diffusion and not
quantum mechanics.

\section{Non-dissipative diffusion}

In order to construct a complex wave function in addition to $\rho$ we require
a second independent degree of freedom that will be identified with the phase
of the wave function. The problem is that the externally prescribed potential
$\phi$ is not an independent degree of freedom. The solution is to change the
constraint by promoting the potential $\phi$, or equivalently $\Phi$ in
eq.(\ref{curr}), to a fully dynamical degree of freedom. This is achieved by
readjusting the potential $\phi$ at each time step in response to the evolving
$\rho$. The appropriate constraint arises from imposing that the potential
$\phi$ be updated in such a way that a certain functional, that we will later
call \textquotedblleft energy\textquotedblright, remains constant. Thus the
dynamics consists in the coupled non-dissipative evolution of $\rho(x,t)$ and
$\Phi(x,t)$.

In the standard approaches to dynamics the conservation of energy is derived
from an action principle plus symmetry under time translations. This approach
is not open to us because we do not have access to an action principle. In
order to define equations of joint evolution for $\rho$ and $\Phi$ we must
identify the relevant constraints. Accordingly, the logic of our derivation
runs in the opposite direction: we first identify the conservation of an
energy and the invariance of the expression for energy under time translations
as the pieces of information that are relevant to our inferences and then we
derive Hamilton's equations and its associated action principle.

\paragraph*{The ensemble Hamiltonian}

For the quantum systems that interest us, the energy functional that codifies
the correct constraint is of the form (\ref{Hamiltonian a}). We therefore
impose that, irrespective of the initial conditions, the potential $\phi$ will
be updated in such a way that the functional $H[\rho,\Phi]$ in
(\ref{Hamiltonian a}) is always conserved,
\begin{equation}
\frac{dH}{dt}=\int dx\,\left[  \frac{\delta H}{\delta\Phi}\partial_{t}%
\Phi+\frac{\delta H}{\delta\rho}\partial_{t}\rho\right]  =0~.
\end{equation}
Using eq.(\ref{Hamilton a}) we get
\begin{equation}
\frac{dH}{dt}=\int dx\,\left[  \partial_{t}\Phi+\frac{\delta H}{\delta\rho
}\right]  \partial_{t}\rho=0~. \label{dHdt}%
\end{equation}
We require that $dH/dt=0$ hold for arbitrary choices of the initial values of
$\rho$ and $\Phi$. Using eq.(\ref{FP c}) we see that this amounts to imposing
$dH/dt=0$ for arbitrary choices of $\partial_{t}\rho$. Therefore the factor in
brackets in eq.(\ref{dHdt}) must vanish at the initial $t_{0}$. But $t_{0}$ is
arbitrary --- any time $t$ can be taken as the initial time for evolution into
the future. Therefore the requirement that $H$ be conserved for arbitrary
initial conditions amounts to imposing that
\begin{equation}
\partial_{t}\Phi=-\frac{\delta H}{\delta\rho} \label{Hamilton b}%
\end{equation}
for all values of $t$. At this point we recognize that eqs.(\ref{Hamilton a})
and (\ref{Hamilton b}) have the form of a canonically conjugate pair of
Hamilton's equations with the conserved functional $H[\rho,\Phi]$ in
(\ref{Hamiltonian a}) playing the role of the Hamiltonian.

\noindent\textbf{Remark:} Note that one can start talking about a Hamiltonian
only \emph{after} a considerable amount of the ED formalism is in place. In
particular, first one must introduce the notion of time, and then one can show
that a suitable choice of constraints leads to a Hamiltonian dynamics.

\paragraph*{The action, Poisson brackets, etc.}

The field $\rho$ is a generalized coordinate and $\Phi$ is its canonical
momentum. Eq.(\ref{Hamilton b}) leads to a generalized Hamilton-Jacobi
equation,%
\begin{equation}
\partial_{t}\Phi=-\frac{1}{2}m^{AB}\partial_{A}\Phi\partial_{B}\Phi
-\frac{\delta F}{\delta\rho}~. \label{HJ}%
\end{equation}
It is easy to check that Hamilton's equations, (\ref{Hamilton a}) and
(\ref{Hamilton b}), can be derived from an action principle
\begin{equation}
\delta A=0\quad\text{where}\quad A[\rho,\Phi]=\int dt\left(  \int dx\,\Phi
\dot{\rho}-H[\rho,\Phi]\right)  ~.
\end{equation}
The time evolution of any arbitrary function $f[\rho,\Phi]$ is given by a
Poisson bracket,
\begin{equation}
\frac{d}{dt}f[\rho,\Phi]=\int dx\left[  \frac{\delta f}{\delta\rho}%
\frac{\delta H}{\delta\Phi}-\frac{\delta f}{\delta\Phi}\frac{\delta H}%
{\delta\rho}\right]  =\left\{  f,H\right\}  \,,
\end{equation}
so that $H$ is the generator of time evolution. Similarly one can check that
$P_{A}=\int dx\rho\partial_{A}\Phi$ is a kind of momentum\ --- it is the
generator of translations in configuration space.

\paragraph*{A Schr\"{o}dinger-like equation}

Given $\rho$ and $\Phi$ we can always combine them into a single complex
function,
\begin{equation}
\Psi_{k}=\rho^{1/2}\exp(ik\Phi/\eta)\,,~ \label{psi k}%
\end{equation}
where $k$ is some arbitrary positive constant the choice of which will be
discussed below. The two coupled equations (\ref{Hamilton a}) and
(\ref{Hamilton b}) can then be written as a single complex
Schr\"{o}dinger-like\ equation,
\begin{equation}
i\frac{\eta}{k}\partial_{t}\Psi_{k}=-\frac{1}{2}\frac{\eta^{2}}{k^{2}}%
\,m^{AB}\partial_{A}\partial_{B}\Psi_{k}+\frac{1}{2}\frac{\eta^{2}}{k^{2}%
}\,m^{AB}\frac{\partial_{A}\partial_{B}|\Psi_{k}|}{|\Psi_{k}|}\Psi_{k}%
+\frac{\delta F}{\delta\rho}\Psi_{k}~. \label{sch a}%
\end{equation}

\section{Information geometry again: the Schr\"{o}dinger equation}

Next we discuss the choice of the functional $F[\rho]$. Let us first recall
the definition of the Fisher information matrix. Consider the family of
distributions $\rho(x|\theta)$ that are generated from a distribution
$\rho(x)$ by pure translations by a vector $\theta^{A}$, $\rho(x|\theta
)=\rho(x-\theta)$. The extent to which $\rho(x|\theta)$ can be distinguished
from the slightly displaced $\rho(x|\theta+d\theta)$ or, equivalently, the
information distance between $\theta^{A}$ and $\theta^{A}+d\theta^{A}$, is
given by $d\ell^{2}=g_{AB}d\theta^{A}d\theta^{B}$ where%
\begin{equation}
g_{AB}(\theta)=\int d^{3N}x\frac{1}{\rho(x-\theta)}\frac{\partial\rho
(x-\theta)}{\partial\theta^{A}}\frac{\partial\rho(x-\theta)}{\partial
\theta^{B}}~.
\end{equation}
Changing variables $x-\theta\rightarrow x$ yields the Fisher information
matrix,%
\begin{equation}
g_{AB}(\theta)=\int d^{3N}x\frac{1}{\rho(x)}\frac{\partial\rho(x)}{\partial
x^{A}}\frac{\partial\rho(x)}{\partial x^{B}}=I_{AB}[\rho]~. \label{Fisher}%
\end{equation}

\paragraph*{The functional $F[\rho]$}

The simplest choice of functional $F[\rho]$ is linear in $\rho$, $F[\rho]=\int
d^{3N}x\,\rho V$, where $V(x)$ is some function that will be recognized as the
familiar scalar potential. Since ED aims to derive the laws of physics from a
framework for inference it is natural to expect that the Hamiltonian might
also contain terms that are of a purely informational nature. We have
identified two such tensors: one is the information metric of configuration
space $\gamma_{AB}\propto m_{AB}$, the other is $I_{AB}[\rho]$. The simplest
nontrivial scalar that can be constructed from them is the trace $m^{AB}%
I_{AB}$. This suggests%
\begin{equation}
F[\rho]=\xi m^{AB}I_{AB}[\rho]+\int d^{3N}x\,\rho V~,~ \label{QP a}%
\end{equation}
where $\xi>0$ is a constant that regulates the realtive strength of the two
contributions. From eq.(\ref{Fisher}) we see that $m^{AB}I_{AB}$ is a
contribution to the energy such that those states that are more smoothly
spread out tend to have lower energy.\footnote{The term $m^{AB}I_{AB}$ is
sometimes called the \textquotedblleft quantum\textquotedblright\ or the
\textquotedblleft osmotic\textquotedblright\ potential but, given its
epistemic nature, we should refrain from interpreting it as being either a
\textquotedblleft potential\textquotedblright\ or a \textquotedblleft
kinetic\textquotedblright\ energy. The relation between the quantum potential
and the Fisher information was pointed out in \cite{Reginatto 1998}. The case
$\xi<0$ leads to instabilities and is therefore excluded; the case $\xi=0$
leads to a qualitatively different theory and will be discussed elsewhere.}

Substituting eq.(\ref{QP a}) into (\ref{sch a}) gives a non-linear
Schr\"{o}dinger equation,%
\begin{equation}
i\frac{\eta}{k}\partial_{t}\Psi_{k}=-\frac{\eta^{2}}{2k^{2}}m^{AB}\partial
_{A}\partial_{B}\Psi_{k}+\left(  \frac{\eta^{2}}{2k^{2}}-4\xi\right)
m^{AB}\frac{\partial_{A}\partial_{B}|\Psi_{k}|}{|\Psi_{k}|}\Psi_{k}+V\Psi
_{k}~. \label{sch b}%
\end{equation}

\paragraph*{Regraduation}

We can now return to the choice of the arbitrary constant $k$ in $\Psi_{k}$,
eq.(\ref{psi k}). Since the physics is fully described by $\rho$ and $\Phi$
the different choices of $k$ lead to different descriptions of the same theory
and among all these equivalent descriptions it is possible to pick one that is
singled out by being extremely convenient --- a process usually known as
`regraduation'.\footnote{Other notable examples of regraduation include the
Kelvin choice of absolute temperature, the Cox derivation of the sum and
product rule for probabilities, and the derivation of the sum and product
rules for quantum amplitudes.} The optimal choice of $k$, which we denote with
a hat,
\begin{equation}
\hat{k}=(\frac{\eta^{2}}{8\xi})^{1/2}~,
\end{equation}
is such that the non-linear term in eq.(\ref{sch b}) drops out. We then
identify the optimal regraduated $\eta/\hat{k}$ with Planck's constant $\hbar
$,
\begin{equation}
\frac{\eta}{\hat{k}}=(8\xi)^{1/2}=\hbar~,
\end{equation}
and eq.(\ref{sch b}) becomes the linear Schr\"{o}dinger equation,%
\begin{equation}
i\hbar\partial_{t}\Psi=-\frac{\hbar^{2}}{2}m^{AB}\partial_{A}\partial_{B}%
\Psi+V\Psi=%
{\displaystyle\sum\limits_{n}}
\frac{-\hbar^{2}}{2m_{n}}\nabla_{n}^{2}\Psi+V\Psi~, \label{sch c}%
\end{equation}
where the wave function is $\Psi=\rho e^{i\Phi/\hbar}$. The constant
$\xi=\hbar^{2}/8\ $in eq.(\ref{QP a}) turns out to be crucial: it defines the
value of what we call Planck's constant and sets the scale that separates
quantum from classical regimes.

\paragraph*{Discussion}

We conclude that for any positive value of the constant $\xi$ it is always
possible to regraduate $\Psi_{k}$ to a physically equivalent but more
convenient description where the Schr\"{o}dinger equation is linear. From this
entropic perspective the linear superposition principle and the complex
Hilbert spaces are important because they are extremely convenient but not
because they are fundamental. Note also that the linearity of quantum
mechanics is quite robust: once we adopt a non-dissipative Hamiltonian
diffusion, and the information-inspired quantum potential, any value of
$\xi>0$ leads to a linear quantum theory.

The question of whether the Fokker-Planck and the generalized Hamilton-Jacobi
equations, eqs.(\ref{Hamilton a}) and (\ref{Hamilton b}), are fully equivalent
to the Schr\"{o}dinger equation was first raised by Wallstrom in the context
of Nelson's stochastic mechanics and concerns the single- or multi-valuedness
of phases and wave functions. \cite{Wallstrom 1989} Wallstrom objected that
stochastic mechanics will lead to phases $\Phi$ and wave functions $\Psi$ that
are either both multi-valued or both single-valued. Both alternatives are
unsatisfactory: quantum mechanics forbids multi-valued wave functions, while
single-valued phases can exclude physically relevant states (\emph{e.g.},
states with non-zero angular momentum). We will not discuss the Wallstrom's
objection in any detail except to note that it does not arise in the ED
approach described here once particle spin is incorporated into the formalism
(a similar result is valid for the hydrodynamical formalism, as was shown by
Takabayasi \cite{Takabayasi 1983}). Indeed, earlier we briefly mentioned that
the potential $\phi(\vec{x})$ is to be interpreted as an angle. Then
integrating the phase $d\Phi$ over a closed path gives
\begin{equation}%
{\displaystyle\oint}
\vec{\nabla}\Phi\cdot d\vec{\ell}=%
{\displaystyle\oint}
\vec{\nabla}\phi\cdot d\vec{\ell}=2\pi n
\end{equation}
where $n$ is an integer. This is precisely the quantization condition that
leads to full equivalence between ED and the Schr\"{o}dinger equation because
it guarantees that wave functions will remain single-valued even for
multi-valued phases.

\paragraph*{Acknowledgments}

We would like to thank C. Cafaro, N. Caticha, S. DiFranzo, A. Giffin, P.
Goyal, M.J.W. Hall, S. Ipek, D.T. Johnson, K. Knuth, S. Nawaz, C.
Rodr\'{\i}guez, and J. Skilling for many discussions on entropy, inference and
quantum mechanics.


\begin{thebibliography}{9}                                                                                                %


\bibitem {Caticha 2010}A. Caticha, J. Phys. A: Math. Theor.\textbf{ 44},
225303 (2011); arXiv.org/abs/1005.2357.

\bibitem {Caticha 2014}A.Caticha, J. Phys.: Conf. Ser. \textbf{504}, 012009
(2014); arXiv:1403.3822.

\bibitem {Reginatto 2013}M. Reginatto, \textquotedblleft From information to
quanta: a derivation of the geometric formulation of quantum theory from
information geometry\textquotedblright, arXiv:1312.0429.

\bibitem {Caticha 2012}A. Caticha, \emph{Entropic Inference and the
Foundations of Physics} (USP Press, S\~{a}o Paulo, Brazil 2012); online at http://www.albany.edu/physics/ACaticha-EIFP-book.pdf.

\bibitem {Ipek Caticha 2014}S. Ipek and A. Caticha, \textquotedblleft Entropic
Quantization of Scalar Fields\textquotedblright, in these proceedings (2014).

\bibitem {Reginatto 1998}M. Reginatto, Phys. Rev. \textbf{A 58}, 1775 (1998).

\bibitem {Wallstrom 1989}T. C. Wallstrom, Found. Phys. Lett. \textbf{2}, 113
(1989); Phys. Rev. \textbf{A49}, 1613 (1994).

\bibitem {Takabayasi 1983}T. Takabayasi, Prog. Theor. Phys. \textbf{70}, 1 (1983).
\end{thebibliography}
\end{document}